\documentclass[english,keywords,amsmath,amssymb,twocolumn]{revtex4}
\usepackage[T1]{fontenc}
\usepackage[latin1]{inputenc}
\usepackage{braket}
\usepackage{babel}%
\usepackage{graphicx}
\usepackage{color}
\usepackage{bm}
\usepackage{longtable}
\usepackage{amsmath}
\usepackage{amsfonts}
\usepackage{dsfont}
\usepackage{amssymb}
\usepackage{hyperref}
\begin{document}
\title{Local preservation of no-signaling in multiparty $PT$-symmetric evolutions}
\author{Asmita Kumari}
\author{Ujjwal Sen}
\affiliation{Harish-Chandra Research Institute, HBNI, Chhatnag Road, Jhunsi, Allahabad 211 019, India}

\begin{abstract}
Violation of the no-signaling principle may occur in \(PT\)-symmetric evolutions, that is, evolutions that stem from Hamiltonians that are symmetric with respect to parity and time-reversal, of bipartite entangled states. The violation can be avoided by using a non-conventional inner product. We show that even within the formalism that utilizes the conventional Dirac inner product between physical state vectors, it is possible to obtain instances of local preservation of the no-signaling principle for evolutions corresponding to \(PT\)-symmetric non-hermitian Hamiltonians with real eigenvalues, of multiparty entangled states, whose bipartite versions still violate the principle. The phenomenon can be witnessed already by using the 
Greenberger-Horne-Zeilinger state. Interestingly, the generalized W states do not support such a local preservation of no-signaling. 
\end{abstract}


\maketitle
\section{Introduction}
 
 A hermitian Hamiltonian and a corresponding unitary dynamics form one of the basic postulates of the standard quantum mechanical description of physical systems \cite{balo-kothai-tomar-desh}. 
 However, like all the other postulates of quantum mechanics, the postulate governing quantum dynamics has also been ``tweaked'', and one of them is the proposal to use evolutions corresponding to non-hermitian Hamiltonians with real eigenvalues that are symmetric with respect to parity and time-reversal \cite{bender98}. Such ``\(PT\)-symmetric'' Hamiltonians with real eigenvalues does not lead to unitary evolutions via the usual exponentiation, and their eigenvectors corresponding to distinct eigenvalues are in general not orthogonal. The exponentiated \(PT\)-symmetric Hamiltonian, when used as the evolution operator, does not preserve the normalization of the input state, and the usual escape is to renormalize the output, before being used for further study. See \cite{eijey, bender02,brody12, current-nei1, current-nei2} and references therein for further properties and subtleties.
 

$PT$-symmetric theory has found use in several areas, although such treatments remain contentious. See 
e.g. \cite{ruter,chang,graefe,kreibich,deffner,
bender17,gunther,lee, bro16, japa, amar-balar-kichhu-chhilo-na, nagarik-klantite,pati15,pati19,pati14} and references therein. In particular, it has recently been shown 
in
\cite{lee}  that local operation of the evolution operator, on an entangled shared state \cite{HHHH}, corresponding to a $PT$-symmetric Hamiltonian violates the no-signaling principle. The question has been re-addressed, and in particular, in Refs. \cite{bro16, japa}, it has been shown that an altered inner product of the associated state space can lead to a resurrection of the no-signaling principle. 

The no-signaling principle states that instantaneous communication of discernable information at a distance is not possible. 
The principle has been a guiding one in physical theories, and it is natural to see its status in a \(PT\)-symmetric physical theory. It is intriguing to note here that a general probabilistic description of a physical system allows stronger correlations than are allowed by quantum mechanics, and in particular, can violate a Bell inequality \cite{bell} stronger than any quantum correlated state. See \cite{pope}, and references thereto. 


Multiparty entanglement is known to lead to phenomena and provide utility that bipartite entanglement is not able or not efficient for \cite{HHHH}. It is therefore natural to investigate the status of the no-signaling principle within a \(PT\)-symmetric physical theory in a multiparty setting. We find that while remaining within the sphere of the traditional inner product of Hilbert spaces, there exists \(PT\)-symmetric non-hermitian Hamiltonians 
with real eigenvalues that 
satisfies the no-signaling principle in a multiparty setting for a given multiparty entangled state, while the same state when considered in a bipartite setting leads to signaling for the same non-hermitian Hamiltonian. This can be seen already for the  Greenberger-Horne-Zeilinger (GHZ) \cite{GHZ}. The generalized W states in the computational basis, however, do not support the phenomenon of local preservation of no-signaling \cite{swapna-madhur-mohe}.


The rest of the 
paper is organized as follows. In Section \ref{dui}, we discuss the most general two-dimensional $PT$-symmetric Hamiltonian.
The result in  \cite{lee} for bipartite systems has been reproduced in Section \ref{tin}. The multiparty setting is taken up in Section \ref{char}. 
We present a conclusion 
Section \ref{pienc}.  

\section{Hamiltonian of a $PT$-symmetric spin-1/2 system}
\label{dui}
 The most general $PT$-symmetric Hamiltonian of a system described on a two-dimensional complex Hilbert space is
 \(J H_{PT}\), where \(H_{PT}\) given by \cite{eijey}
\begin{eqnarray}
\label{evo} 
H_{PT}=\left(
\begin{array}{cc}
  r+t \cos\xi-i s \sin\xi & i s \cos\xi+t \sin\xi \\
 i s \cos\xi+t \sin\xi & r-t \cos\xi+i s \sin\xi \\
\end{array}
\right),\nonumber\\
\end{eqnarray}
where $r, s, t, \xi$ are real parameters. \(J\) is a nonzero real number that has the unit of ``energy'', so that the other parameters are dimensionless. The operators corresponding to parity (\(P\)) and time-reversal (\(T\)) are respectively 
\begin{eqnarray}
\label{phaTkabajir-deshe} 
P=\left(
\begin{array}{cc}
  \cos\tilde{\phi} & \sin\tilde{\phi} \\
\sin\tilde{\phi} & -\cos\tilde{\phi} \\
\end{array}
\right), \nonumber
\end{eqnarray}
with a real \(\tilde{\phi}\),
and complex conjugation in the computational basis, which is the eigenbasis of the Pauli \(\sigma_z\) operator \cite{eijey}. 
Here we are concerned with the family of $H_{PT}$ having real eigenvalues, for which we need \(s^2 \leq t^2\). Substituting $\sin \alpha = s/t$,
the 
real eigenvalues of $J H_{PT}$ are $E_{\pm} = J(r \pm t \cos\alpha)$, with the  corresponding 
eigenstates being 
\begin{eqnarray} 
|E_{\pm} (\alpha) \rangle = \frac{1}{N_1}\left(
\begin{array}{c}
  \sin{\alpha}\cos{\xi/2}- i (1\mp \cos\alpha)\cos{\xi/2}  \\
 \sin{\alpha}\cos{\xi/2}+ i (1\mp \cos\alpha)\cos{\xi/2} \\
\end{array}
\right), \nonumber \\ 
\end{eqnarray}
where $N_1=\sqrt{2(1\mp \cos\alpha)\cos\alpha}$ is the normalization constant. It should be noted that unlike eigenstates of a hermitian operator, $|E_{+} (\alpha) \rangle$ and $|E_{-} (\alpha) \rangle$ are nonorthogonal states, and they coincide with each other at $\alpha = \pm \pi /2$.

\section{Bipartite entangled state and violation of no-signaling principle}
\label{tin}

It has been shown in Ref. \cite{lee} that a bipartite entangled state of a system consisting of two spin-1/2 systems described by local \(PT\)-symmetric Hamiltonians can be used to violate the no-signaling principle. We present that analysis below for the evolution corresponding to the most general \(PT\)-symmetric Hamiltonian acting on one of the two spins-1/2.

 Let Alice and Bob be in two space-like separated locations, sharing the maximally entangled state given by
\begin{eqnarray}
|\psi \rangle_{AB} = \frac{1}{\sqrt{2}}(|00 \rangle + |11 \rangle),
\end{eqnarray}
where \(|0\rangle\) and \(|1\rangle\) are eigenstates of the Pauli \(\sigma_z\) operator.
Let the corresponding density matrix be \(\rho_{AB}\), so that  $\rho_{AB} = |\psi \rangle_{AB} \langle \psi |$. Then under any unitary dynamics by Alice, the reduced density matrix of Bob is 
\begin{eqnarray}
\label{ns1}
 \rho_{B} =\frac{1}{2}\left(
\begin{array}{cc}
  1 & 0 \\
 0 & 1 \\
\end{array}
\right), 
\end{eqnarray}
before as well as after the evolution.
This is 
the no-signaling condition for the maximally entangled state within quantum evolutions, which says that the probability distribution of outcomes of Bob is unaffected by Alice's choice of quantum operations. Indeed, the no-signaling holds 
even if Alice chooses to perform a unitary operation on an augmented system on her side. 

 If Alice evolves her part of the shared maximally entangled state by using the $PT$-symmetric Hamiltonian given in Eq.(\ref{evo}),
 then the composite state reduces to
\begin{eqnarray}
|\psi \rangle^U_{AB} &=& U \otimes \mathbb{I} \left(\frac{1}{\sqrt{2}}\left(|00 \rangle + |11 \rangle\right)\right) \\ \nonumber
&=& \frac{1}{\sqrt{2}}\left(U|0 \rangle \otimes |0 \rangle+ U|1 \rangle \otimes |1 \rangle\right), 
\end{eqnarray}
where the non-unitary evolution $U =e^{-i H_{PT} \tau}$, where \(\tau = J \tau{'}/\hbar\), with \(\tau{'}\) playing the role of ``time''. The operator \(\mathbb{I}\) denotes the identity operator on the two-dimensional complex Hilbert space, \(\mathbb{C}^2\). The composite density matrix becomes
\begin{eqnarray}
\rho_{AB}^{U} = U \otimes \mathbb{I} \rho_{AB} U^{\dagger} \otimes  \mathbb{I}.
\end{eqnarray}
 Now by taking partial trace of Alice's part, the reduced state of Bob, after renormalization, is obtained as
\begin{eqnarray}
\rho_{B}^{U} &=& \mbox{Tr}_A[\rho_{AB}^{U}] = \frac{1}{N_2} \left(
\begin{array}{cc}
  b_1 & b_4 \\
 b_3 & b_2 \\
\end{array}
\right), 
\end{eqnarray}
where $N_2 = 2(2 \sec ^2\alpha  \sin ^2t_1+\cos (2 t_1))$, and
\begin{eqnarray}
\label{jene-shune-bish}
\nonumber
b_1 &=&  
-\tan \alpha  \sin (2 t_1) \sin \xi  
+2 \sec ^2\alpha  \sin ^2 t_1+\cos (2 t_1), \\ \nonumber
b_2 &=& 
\tan \alpha  \sin (2 t_1) \sin \xi 
+ 2 \sec ^2 \alpha  \sin ^2 t_1 +\cos (2 t_1), \\ \nonumber
b_3 &=& 
2 \tan \alpha  \sin t_1 \left(\cos t_1 \cos \xi 
+ i \sec \alpha  \sin t_1 \right), \\ 
b_4 &=& 
2 \tan \alpha  \sin t_1 \left(\cos t_1 \cos \xi 
- i \sec \alpha  \sin t_1 \right).
\end{eqnarray}
Here $t_1 = t \tau \cos\alpha$.

The no-signaling principle is satisfied if  $\rho_{B}^U = \mathbb{I}/2$. However, that happens only if \(\alpha =0\) or \(t_1 =0\). \(\alpha =0\) leads to a hermitian \(H_{PT}\). \(t_1 = 0\) implies \(t=0\), which again leads to a hermitian \(H_{PT}\) as then \(s=0\), or \(\tau =0\), which means that the evolution did not happen, or \(\alpha = \pm \pi/2\), which is the case when the eigenvectors of the \(PT\)-symmetric Hamiltonian coincide. So, except for the case when \(\alpha = \pm \pi/2\), all other \(PT\)-symmetric non-hermitian Hamiltonians with real eigenvalues lead to signaling of information by using the pre-shared maximally entangled state, consistent  
%
with the result obtained in \cite{lee}.

\section{The multisite scenario}
\label{char}

The beautiful simplicity of the case of two spins-1/2 with respect to \(PT\)-symmetric evolutions and the no-signaling principle, becomes far richer already for three spins-1/2.
We now suppose that Alice, Bob, and Charu pre-share a state, possibly entangled, of three spins-1/2, and Alice applies a \(PT\)-symmetric ``spin'' on her spin. Application of a unitary evolution by Alice will of course not lead to any signaling of information to Bob and Charu. We are therefore interested in situations where Alice applies the evolution corresponding to a \(PT\)-symmetric non-hermitian Hamiltonian with real eigenvalues. If Bob and Charu are together at the same location, then the analysis of the preceding section applies,  and there is always (except for the symmetry-breaking points at \(\alpha = \pm \pi/2\)) a violation of the no-signaling principle for a state that is maximally entangled in the Alice versus Bob-Charu bipartition. However, if Bob and Charu are far apart, and all the three observers are mutually in space-like separated locations, then we show below that
the no-signaling principle can be satisfied even for some non-hermitian \(PT\)-symmetric Hamiltonians with real eigenvalues with a state that is still maximally entangled in the Alice versus Bob-Charu bipartition. 
Actually, we consider below also cases where the Alice to Bob-Charu bipartition is non-maximally entangled.



\subsection{When the three parties share a generalized GHZ state}
If the state shared between Alice, Bob, and Charu is the generalized GHZ state given by
\begin{eqnarray}
|\psi_1 \rangle_{ABC_{G}} = \cos\theta|000 \rangle + \sin\theta|111 \rangle,
\end{eqnarray}
the corresponding density matrix is  $\rho_{ABC_{G}} = |\psi_1 \rangle_{ABC_{G}} \langle \psi_1 |$, and the joint reduced state of Bob and Charu obtained after taking the partial trace on Alice's part is
\begin{eqnarray}
\rho_{BC_{G}} = \mbox{Tr}_{A}[\rho_{ABC_{G}}]=\left(
\begin{array}{cccc}
 \cos ^2\theta  & 0 & 0 & 0 \\
 0 & 0 & 0 & 0 \\
 0 & 0 & 0 & 0 \\
 0 & 0 & 0 & \sin ^2\theta  \\
\end{array}
\right).
\end{eqnarray}
The reduced states of Bob and Charu are
\begin{eqnarray}
\label{bcw}
\rho_{B_{G}}= \mbox{Tr}_{AB}[\rho_{ABC_{G}}]=\left(
\begin{array}{cc}
\cos ^2\theta  & 0 \\
 0 & \sin ^2\theta  \\
\end{array}
\right) = \rho_{C_{G}}.
\end{eqnarray}
Now let us assume that Alice evolves her part 
using the $PT$-symmetric Hamiltonian $J H_{PT}$ 
($U = e^{-i H_{PT} \tau} $) 
given by Eq.(\ref{evo}), while 
Bob and Charu do not evolve their parts. 
Then the composite density $\rho_{ABC_{G}}$ at time $\tau{'}$ becomes
\begin{eqnarray}
\rho_{ABC_{G}}^U = U \otimes \mathbb{I} \otimes \mathbb{I}(\rho_{ABC_{G}}) U^{\dagger} \otimes  \mathbb{I}\otimes \mathbb{I}, 
\end{eqnarray}
where a renormalization is implicitly assumed.
The joint reduced state of Bob and Charu after Alice's application of \(U\) and after renormalization is
\begin{eqnarray}
\nonumber
\rho^U_{BC_{G}} = \mbox{Tr}_{A}[\rho^U_{ABC_{G}}]=\frac{1}{N_3}\left(
\begin{array}{cccc}
2 b_1 \cos^2 \theta   & 0 & 0 &  b_4 \sin (2 \theta)   \\
 0 & 0 & 0 & 0 \\
 0 & 0 & 0 & 0 \\
  b_3 \sin (2 \theta)  & 0 & 0 & 2 b_2 \sin^2 \theta   \\ 
\end{array}
\right),
\end{eqnarray}
and the density matrices of Bob and Charu are given by 
\begin{eqnarray}
\nonumber
\rho^{U}_{B_{G}}= \mbox{Tr}_{AC}[\rho_{ABC_{G}}^U]= \frac{1}{N_3}\left(
\begin{array}{cc}
2 b_1 \cos^2 \theta  & 0 \\
 0 & 2 b_2 \sin^2 \theta  \\
\end{array}
\right) = \rho^{U}_{C_{G}},
\end{eqnarray}
 where $N_3 = 2(\sec ^2 \alpha - \tan \alpha  (\tan \alpha  \cos (2 t_1)+\cos (2 \theta ) \sin (2 t_1) \sin \xi ))$ and $t_1 = t \tau \cos \alpha$.
 Here, \(b_1\), \(b_2\), \(b_3\), and \(b_4\) are given by (\ref{jene-shune-bish}).
 
It may be noted that the state \(|\psi_1 \rangle_{ABC_{G}}\) is equivalent to the state 
\(\cos \theta |00\rangle + \sin\theta |11\rangle\) in the bipartite scenario, when Bob and Charu are at the same location. And for \(\theta=\pi/4\), the state the maximally entangled state considered in the preceding section.

Comparing $\rho^{U}_{BC_{G}}$ with $\rho_{BC_{G}} $, we find that the no-signaling principle is satisfied ($\rho^{U}_{BC_{G}}  = \rho_{BC_{G}} $) if and only if $t_1 = 0$ or $\alpha = 0 $, consistent with the calculations of the preceding section. As we have discussed in the preceding section, $t_1 = t \tau \cos \alpha = 0$ or $\alpha = 0 $ convert the $PT$-symmetric Hamiltonian to a hermitian one, except when \(\alpha = \pm \pi/2\), where \(H_{PT}\) has equal eigenvectors.
Hence we can claim that for the scenario where Bob and Charu are at the same location, 
the no-signaling principle is violated if the local operation on Alice's part is implemented by resorting to a 
$PT$-symmetric non-hermitian Hamiltonian with real eigenvalues. However, if Bob and Charu are at separate locations, we need to compare the post-evolution reduced density matrices of Bob and Charu, viz.   $\rho^{U}_{B_{G}}$ and $\rho^{U}_{C_{G}}$ respectively, with the initial ones, viz.   $\rho_{B_{G}}$  and $\rho_{C_{G}}$ respectively. 

We find that again there is violation of the no-signaling principle, so that 
$\rho^{U}_{B_{G}} \ne \rho_{B_{G}}$ and $\rho^{U}_{C_{G}} \ne \rho_{C_{G}}$, 
except for the surface of the parameter space given by \(\xi =0\). On the surface \(\xi =0\), 
we find that for arbitrary \(\theta\), and for arbitrary \(r\), \(s\), \(t\), \(J\), \(\tau{'}\), the no-signaling condition holds true for the local systems, i.e., 
$\rho^{U}_{B_{G}} = \rho_{B_{G}}$ and $\rho^{U}_{C_{G}} = \rho_{C_{G}}$.
 Here, \(\theta\), \(r\), \(s\), \(t\), \(J\), \(\tau{'}\) are all real, \(J\) and \(\tau{'}\) are also nonzero, and \(s^2 < t^2\).

We therefore find that there can be multiparty states for which nontrivial non-hermitian \(PT\)-symmetric Hamiltonians with real eigenvalues that can give rise to violation of no-signaling when some of the parties sharing the multiparty state are together, while the no-signaling holds when they are all far apart. We term this as ``local preservation of no-signaling''. It is to be noted that there are ample \(PT\)-symmetric Hamiltonians with real eigenvalues which do violate the no-signaling principle for the same shared states. 


While we have used non-unitary operators to perform time evolution of the system, if we consider the usual quantum measurements to distinguish between the states \(\rho^U_{BC_G}\) and \(\rho_{BC_G}\)
before and after the evolution at Alice's end, then we will obtain a quantification of the amount of violation of the no-signaling principle, for the given shared state. Since a quantum evolution at Alice would have kept the state at Bob and Charu invariant, the same quantity is a measure of the amount by which we have violated quantum mechanics for the given shared state.

Consider a source creating either of a pair of quantum states, 
\(\rho_1\) and \(\rho_2\),  with equal probabilities. The maximum probability of ``minimum error discrimination'' between the states by the best quantum measurement strategy is given by the 
Helstrom quantity \cite{madhabi}
\begin{eqnarray}
P = \frac{1}{2}+\frac{1}{4}||\rho_{1} - \rho_{2}||_1. 
\end{eqnarray}
We use this probability as the distinguishing quantity between the pre- and post-measurement states.



In the case of a generalized $GHZ$ state, the best success probability of discriminating between $\rho_{BC_G}$ and $\rho^U_{BC_G}$ is given by
\begin{eqnarray}
\label{ps11}
P_{GHZ} = \frac{1}{2}+\frac{1}{4}||\rho^U_{BC_G} - \rho_{BC_G}||_1.
\end{eqnarray}
The four eigenvalues of $\rho_{d_{1}} = \rho^{U}_{BC_G} - \rho_{BC_G}$ are $\lambda_1 = 0$, $\lambda_2 = 0$,
\begin{eqnarray}
\nonumber \lambda_3= \frac{1}{4 N_4}\bigg[ \cos \alpha  \sin (2 \theta ) \sin t_1 \bigg(-4 \cos ^2 \alpha  \cos (4 \theta ) \sin ^2 \xi \\ \nonumber    +\cos (2 \alpha ) (\cos (2 \xi )+3) - \cos (2 t_1) \{4 \cos ^2 \alpha  \cos (4 \theta ) \sin ^2 \xi \\ \nonumber  -\cos (2 \alpha ) (\cos (2 \xi )+3)-\cos (2 \xi )+5\} \\ \nonumber +\cos (2 \xi )+11  \bigg)^{-1/2}    \bigg],
\end{eqnarray}
and $\lambda_4 = -\lambda_3$, where $N_4 = \sin \alpha  \cos \alpha  \cos (2 \theta ) \sin (2 t_1) \sin \xi +\sin^2 \alpha  \cos (2 t_1)-1$.
Now Eq.~(\ref{ps11}) can now be rewritten as
\begin{equation}
P_{GHZ} = \frac{1}{2}\left(1 + \left|\lambda_3\right|\right)
\end{equation}
%

\begin{figure}[t!]
\centering
\includegraphics[width=1\linewidth]{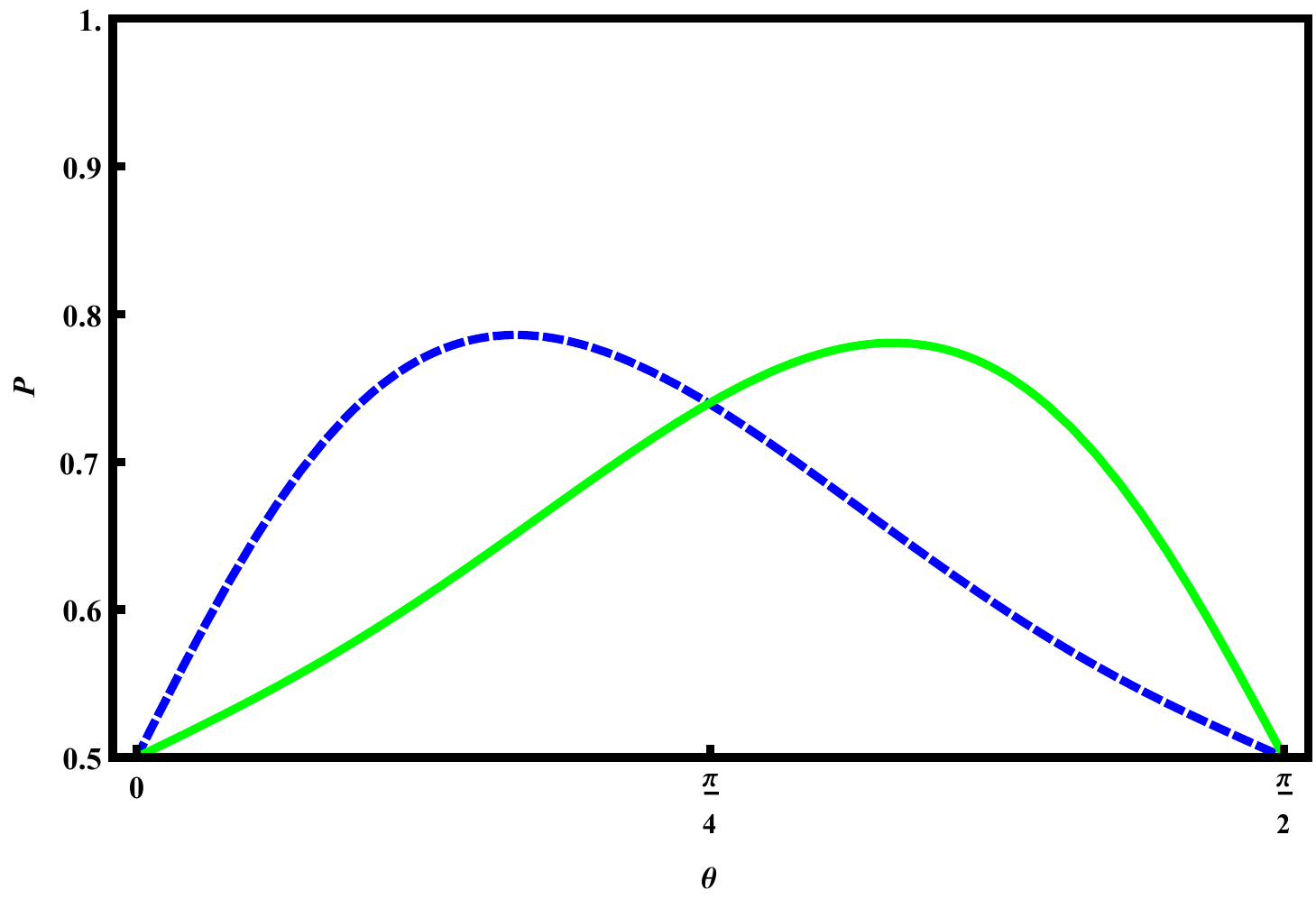}
\caption{Maximum success probability of minimum error discrimination between states before and after evolution with \(PT\)-symmetric Hamiltonians. For the blue dashed profile, the horizontal axis is the parameter \(\theta\) of the generalized GHZ state. The vertical axis is the maximal probability of minimum error discrimination between the joint states of Bob and Charu before and after evolution by Alice with the Hamiltonian \(JH_{PT}\), when the three parties share a generalized GHZ state, where \(t_1 = \pi/18\), \(\xi = \pi/2\), and \(\alpha = 9\pi/20\). The highest probability obtained in the entire parameter space (of the shared state and the evolution) lies on this profile at \(\theta \approx 0.519\). 
The green continuous curve is for the case when the three-party shared state is the generalized W state, and where
\(\alpha = \frac{2 \pi }{5}\), \(\xi = \frac{\pi }{2}\) , \(t_1 = 0.3479\), and \(\phi = \frac{\pi }{4}\). Again, the highest probability obtained in the entire parameter space lies on this slice, and is at $\theta \approx \pi/3$.
We have performed the same analysis for more general states, and much higher probabilities can  be obtained in such cases. 
The horizontal axis is measured in radians, while the vertical one is dimensionless, for 
both the curves.}
\label{fig:1}
\end{figure}

\begin{figure}[t!]
\centering
\includegraphics[width=1\linewidth]{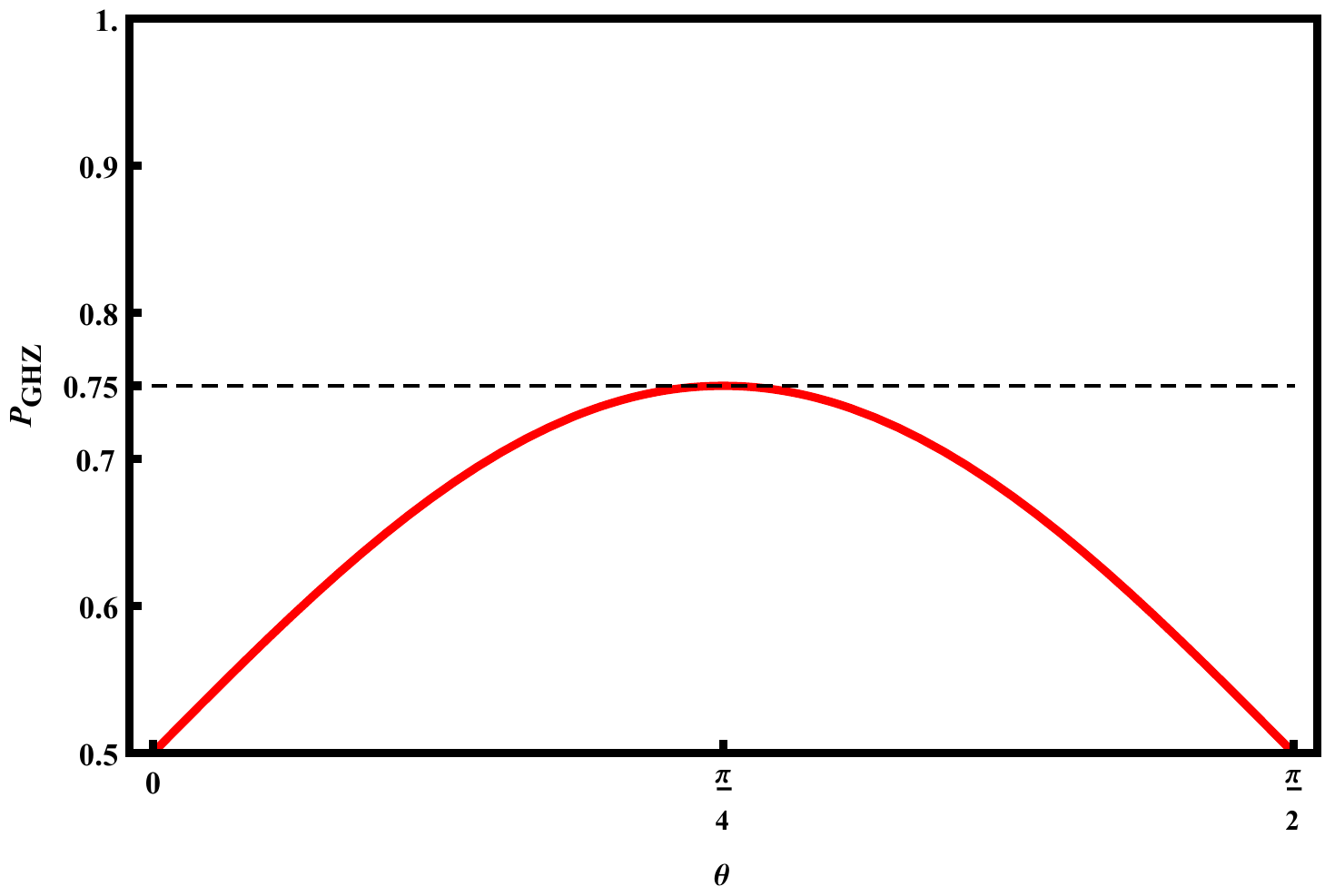}
\caption{How much violation of no-signaling is present when it is avoided locally for generalized GHZ states? The horizontal axis represents the parameter \(\theta\) of the generalized GHZ state. The vertical axis represents the maximal probability of success for minimum error discrimination between the joint states at Bob and Charu before and after an evolution by Alice with a \(PT\)-symmetric non-hermitian Hamiltonian with real eigenvalues, maximized over all parameters of the Hamiltonian but on the surface \(\xi=0\), and when the shared state is a generalized GHZ state. The horizontal axis is in radians, while the vertical one is dimensionless.}
\label{madhup}
\end{figure}

We found that the best success probability of discriminating between $\rho_{BC_G}$ and $\rho^U_{BC_G}$ is $ \approx 0.786$, obtained at $t_1 \approx \pi/18\), \(\theta \approx 0.519\), \(\xi \approx \pi/2 $, and 
\(\alpha  \approx 9 \pi /20 \). The plot of the maximum success probability of discriminating between $\rho_{BC_G}$ and $\rho^U_{BC_G}$ ($P_{GHZ}$) as function of $\theta$ at $t_1 = \pi/18, \xi = \pi/2 $ and $\alpha =9 \pi /20 $ is given in Fig. \ref{fig:1} (blue dashed line). 
One important surface in the parameter space is the \(\xi=0\) one, and on this surface, if we perform an optimization of the minimum error probability of discrimination, for an arbitrary fixed \(\theta\), we obtain a profile as exhibited in Fig. \ref{madhup}.  
In the succeeding subsection, we  discuss the issue of local preservation of the no-signaling principle using generalized W states.

\vspace{1cm}

\subsection{When the three parties share a generalized W state}
If the state shared between Alice, Bob, and Charu is a generalized W state, given by 
\begin{eqnarray}
\nonumber
|\psi_2 \rangle_{ABC_{W}} = \sin  \theta  \cos \phi |001 \rangle + \sin \theta  \sin \phi  |010 \rangle  \\
+ \cos \theta  |100 \rangle, 
\end{eqnarray}
the corresponding density matrix is $\rho_{{ABC}_W} = |\psi_2 \rangle_{ABC_{W}} \langle \psi_2 |$, so that the joint reduced density matrix of Bob and Charu is obtained as
\begin{eqnarray}
\rho_{{BC}_W} = \mbox{Tr}_{A}[\rho_{{ABC}_W}] = \phantom{aaaaaaaaaaaaaaaaaa}\nonumber \\ \nonumber \left(
\begin{array}{cccc}
 \cos ^2\theta  & 0 & 0 & 0 \\
 0 & \cos ^2\phi  \sin ^2\theta  & \cos \phi  \sin ^2\theta  \sin \phi  & 0 \\
 0 & \cos \phi  \sin ^2\theta  \sin \phi  & \sin ^2\theta  \sin ^2\phi  & 0 \\
 0 & 0 & 0 & 0 \\
\end{array}
\right), \nonumber\\ 
\end{eqnarray}
and the reduced density matrices of Bob and Charu are
\begin{eqnarray}
\rho_{B_W}& =& \mbox{Tr}_{AC}[\rho_{{ABC}_W}]\nonumber \\ \nonumber &=&\left(
\begin{array}{cc}
 \sin^2\theta\cos^2\phi + \cos^2\theta & 0 \\
 0 & \sin^2\theta\sin^2\phi \\
\end{array}
\right) \\
\end{eqnarray}
and
\begin{eqnarray}
\rho_{C_W} &=& \nonumber \mbox{Tr}_{AC}[\rho_{{ABC}_W}]\\ \nonumber &=&\left(
\begin{array}{cc}
  \sin^2\theta\sin^2\phi + \cos^2\theta & 0 \\
 0 &  \sin^2\theta\cos^2\phi \\
\end{array}
\right) \\
\end{eqnarray}
respectively.

 If Alice evolves her part using the $PT$-symmetric Hamiltonian $JH_{PT}$ given by Eq.~(\ref{evo}), then the composite system $\rho_{ABC_{W}}$ at time $\tau{'}$ can be written as
\begin{eqnarray}
\rho_{ABC_{W}}^U = U \otimes \mathbb{I} \otimes \mathbb{I}(\rho_{ABC_{W}}) U^{\dagger} \otimes  \mathbb{I}\otimes \mathbb{I}, 
\end{eqnarray}
where a renormalization is implicitly assumed. After the local operation, the joint reduced state of Bob and Charu is
\begin{eqnarray}
\nonumber
&&\rho^U_{{BC}_W} = \mbox{Tr}_{A}[\rho^U_{{ABC}_W}] \\ \nonumber &&=\frac{1}{N_5}\left(
\begin{array}{cccc}
 b_2\cos ^2\theta   &  b_3  w_1    & b_4 w_2   & 0 \\
  b_3 w_2   & b_1 \cos ^2\phi  \sin ^2\theta   & \frac{b_1}{2} \sin( 2\phi)  \sin ^2\theta     & 0 \\
 b_4 w_1 & \frac{b_1}{2} \sin( 2\phi) \sin ^2\theta    & b_1 \sin ^2\theta  \sin ^2\phi   & 0 \\
 0 & 0 & 0 & 0 \\
\end{array}
\right)
\end{eqnarray}
where $w_1 = \sin( 2\theta)  \cos \phi $ and $w_2 = \sin( 2\theta)  \sin \phi $
with $N_5 = \sec ^2\alpha +\tan \alpha  (\cos (2 \theta ) \sin \xi  \sin (2 t_1)-\tan \alpha  \cos (2 t_1))$, and where \(b_1\), \(b_2\), \(b_3\), \(b_4\) are given in (\ref{jene-shune-bish}).
The reduced states of Bob and Charu are
\begin{eqnarray}
\nonumber
\rho^U_{B_W} = \phantom{aaaaaaaaaaaaaaaaaaaaaaaaaaaaaaaaaa}\\ \nonumber \frac{1}{N_5}\left(
\begin{array}{cc}
b_1 \sin ^2 \theta  \cos ^2 \phi   + b_2\cos ^2\theta   &   2 b_3\cos \theta  \cos \phi  \sin \theta   \\
  2 b_4 \cos \theta  \sin \theta  \sin \phi    & b_1 \sin ^2 \theta  \sin ^2\phi   \\
\end{array}
\right) 
\end{eqnarray}
and
\begin{eqnarray}
\nonumber
\rho^U_{C_W} = \phantom{aaaaaaaaaaaaaaaaaaaaaaaaaaaaaaaaaa}\\ \nonumber \frac{1}{N_5} \left(
\begin{array}{cc}
b_1 \sin^2 \theta \sin^2 \phi + b_2 \cos^2 \theta   & 2 b_4 \cos \theta  \sin \theta  \sin \phi  \\
 2 b_3 \cos \theta  \cos \phi  \sin \theta  & b_1 \sin^2 \theta\cos^2 \phi  \\
\end{array}
\right) 
\end{eqnarray} 
respectively. Unlike for the generalized GHZ states, the generalized W states however do not provide a nontrivial non-hermitian Hamiltonian within the \(PT\)-symmetric class for which the no-signaling principle can be saved even locally.  


Even though the no-signaling principle is violated both globally and locally by the generalized W states, the values attained by the probabilities for minimum error discrimination between 
$\rho^U_{BC_W}$  and $\rho_{BC_W} $ are comparable to the generalized GHZ case. The said probability is given by
\begin{eqnarray}
\label{ps2}
P_{W} = \frac{1}{2}+\frac{1}{4}||\rho^U_{{BC}_W} - \rho_{{BC}_W}||_1.
\end{eqnarray}
The four eigenvalues of $\rho_{d_{2}} = \rho^U_{{BC}_W} - \rho_{{BC}_W}$ are $\delta_1 = 0$, $\delta_2 = 0$,
\begin{eqnarray}
\nonumber \delta_3 =  \frac{1}{4 N_6}\bigg[\sec \alpha  \sin (2 \theta ) \sin t_1 \bigg(-2 \sin ^2 \xi  \{\cos (4 \theta ) \phantom{+2}\\ 
\nonumber  -2 \sin ^2(2 \theta ) (\cos (2 \alpha ) 
+2 \cos ^2 \alpha  \cos (2 t_1))\} \\ 
\nonumber +2 \sin (2 \phi ) \{2 \cos (2 \alpha ) \cos ^2 \xi 
+(\cos (2 \alpha )-3) \cos (2 t_1)\}    \\
\nonumber +\cos (2 \xi ) \{\sin (2 \phi ) 
\left(4 \cos ^2\alpha  \cos (2 t_1)+2\right)-1\} \\
\nonumber +10 \sin (2 \phi )+1 \bigg)^{-1/2}    \bigg],
\end{eqnarray}
where $N_6 = \sec ^2 \alpha +\tan \alpha  (\cos (2 \theta ) \sin (\xi ) \sin (2 t_1)-\tan \alpha  \cos (2 t_1))$.
and $\delta_4 = - \delta_3$.
 Therefore, Eq.~(\ref{ps2}) can be rewritten as
\begin{eqnarray}
P_W  = \frac{1}{2}\left( 1 + \left|\delta_3\right|\right).
\end{eqnarray}
The maximum value of this success probability is attained at \(\alpha \approx \frac{2 \pi }{5}\), \(\xi \approx \frac{\pi }{2}\) , \(t_1 \approx 0.3479\), \(\phi \approx \frac{\pi }{4}\) and $\theta \approx \pi/3$, and the corresponding maximum value is approximately 
$0.781$, not far from that in the case of generalized GHZ states. A slice of the multidimensional $P_W$ surface, as a function of \(\theta\)
is plotted in Fig. \ref{fig:1} (green continuous profile).


\subsection{GHZ state with a rotated basis}
Let us now suppose that the state shared between Alice, Bob, and Charu is a GHZ state, but with a rotated basis, viz. 
\begin{eqnarray}
|\psi'_3 \rangle_{ABC_G} = \cos\theta|0'0'0' \rangle + \sin\theta|1'1'1' \rangle,
\end{eqnarray}
where $|0' \rangle =  \cos \frac{x}{2} |0 \rangle + \exp (i y) \sin \frac{x}{2}|1 \rangle$ and $|1' \rangle =  \cos \frac{x}{2} |0 \rangle - \exp (i y) \sin \frac{x}{2}|1 \rangle$.  The corresponding density matrix is $\rho'_{{ABC}_G} = |\psi'_3 \rangle_{ABC_{G}} \langle \psi'_3 |$. 
Then, the joint reduced state of Bob and Charu is 
\begin{eqnarray}
&&\rho'_{BC_G} = \mbox{Tr}_{A}[\rho'_{ABC_G}]= \\ \nonumber && \frac{1}{4} \left(
\begin{array}{cccc}
 \frac{1}{2}g_1 & e^{-i y} g_3 &  e^{-i y} g_3 &  e^{-2 i y} \sin ^2 x \\
 e^{i y} g_3  & \sin ^2 x & \sin ^2 x & -e^{-i y}g_4  \\
  e^{i y} g_3  & \sin ^2 x & \sin ^2 x & - e^{-i y} g_4  \\
 e^{2 i y} \sin ^2 x & - e^{i y} g_4 & - e^{i y} g_4  & \frac{1}{2} g_2  \\
\end{array}
\right),
\end{eqnarray}
where $g_1 = \cos (2 x)+4 \cos x \cos (2 \theta )+3  $ ,  $g_2 = \cos (2 x)-4 \cos x \cos (2 \theta )+3$, $g_3 =(\cos x+\cos (2 \theta ))\sin x  $ and $g_4 =(\cos x-\cos (2 \theta ))\sin x $. The reduced states of Bob and Charu are
\begin{eqnarray}
\nonumber
\rho'_{B_G} &=& \frac{1}{2}\left(
\begin{array}{cc}
 1+\cos x  \cos (2 \theta ) &  e^{-i y} \cos (2 \theta ) \sin x \\
  e^{i y} \cos (2 \theta ) \sin x &  (1-\cos x \cos (2 \theta )) \\
\end{array}
\right) \\   &=& \rho'_{C_G}.
\end{eqnarray}
 
 If Alice evolves her part using the $PT$-symmetric Hamiltonian $JH_{PT}$ given by Eq.~(\ref{evo}), then the joint reduced state of Bob and Charu, after renormalization, is
\begin{eqnarray}
\nonumber
\rho^{'U}_{BC_G} && = Tr_{A}[\rho^{'U}_{ABC_G}]= \frac{1}{8 N_7}\left(
\begin{array}{cccc}
g_5 & g_8 & g_8 & 2 g_{11}\\
	8 g_6 & g_9 & g_9  & 8 g_{12} \\
	8 g_6 & g_9 & g_9 & 8 g_{12}  \\
	2 g_7 & 8 g_{10} & 8 g_{10} & g_{13} \\
\end{array}
\right),
\\
\end{eqnarray}
where
\begin{eqnarray}
\nonumber
g_5 &=& 8 b_5 \cos ^2\theta  \cos ^4\frac{x}{2}+8 b_6 \sin ^2 \theta  \sin ^4 \frac{x}{2} \\ \nonumber &+&(b_7+b_8) \sin (2 \theta ) \sin ^2 x, \\ \nonumber 
g_6 &=& \frac{1}{2}e^{i y} \sin x \bigg[\cos \theta  \cos ^2\frac{x}{2} (b_5 \cos \theta \\ \nonumber &-& b_7 \sin \theta )+\sin \theta  \sin ^2 \frac{x}{2} (b_8 \cos \theta -b_6 \sin \theta )\bigg],\\ \nonumber
g_7 &=&e^{2 i y} \bigg[b_5 \cos ^2 \theta  \sin ^2 x +b_6 \sin ^2 \theta  \sin^2 x \\ \nonumber &+&4 \sin \theta  \cos \theta  \big(b_7 \cos ^4 \frac{x}{2}+b_8 \sin ^4 \frac{x}{2} \big)\bigg], \\ \nonumber
g_8 &=& e^{-i y} \sin x \bigg[\cos (2 \theta ) ((b_5-b_6) \cos x +b_5+b_6)\\ \nonumber&+&  b_5 \cos x +b_5+b_6 \cos x-b_6  -\sin (2 \theta ) ((b_7 \\ \nonumber &+&b_8) \cos x-b_7+b_8) \bigg],\\ 
\nonumber
g_9 &=& \sin ^2 x \bigg[(b_5-b_6) \cos (2 \theta )+b_5+b_6\\ \nonumber &-&(b_7+b_8) \sin (2 \theta ) \bigg],
\end{eqnarray}
\begin{eqnarray}
\nonumber
g_{10} &=& \frac{1}{2}e^{i y} \sin x  \bigg[b_5 \cos ^2 \theta  \sin ^2 \frac{x}{2}\\ \nonumber &-&b_6 \sin ^2 \theta  \cos ^2 \frac{x}{2} +\frac{1}{4} \sin (2 \theta ) ((b_7+b_8) \cos x\\ \nonumber &+&b_7-b_8)\bigg], \\ \nonumber
g_{11} &=&  e^{-2 i y} \bigg[b_5 \cos ^2 \theta  \sin ^2 x +b_6 \sin ^2 \theta  \sin ^2 x \\ \nonumber &+&4 \sin \theta  \cos \theta  \big(b_7 \sin ^4 \frac{x}{2} +b_8 \cos ^4 \frac{x}{2} \big)\bigg],\\ \nonumber 
g_{12} &=& \frac{1}{2} e^{-i y} \sin x  \bigg[b_5 \cos ^2 \theta  \sin ^2 \frac{x}{2}\\ \nonumber &-&b_6 \sin ^2 \theta  \cos ^2 \frac{x}{2} +\frac{1}{4} \sin (2 \theta ) ((b_7+b_8) \cos x \\ \nonumber &-&b_7+b_8)\bigg],\\
%
%
\nonumber
g_{13} &=& 8 b_5 \cos ^2 \theta  \sin ^4 \frac{x}{2} +8 b_6 \sin ^2 \theta  \cos ^4 \frac{x}{2}\\ \nonumber &+& (b_7+b_8) \sin (2 \theta ) \sin ^2x. 
\end{eqnarray}
Here, $b_5, b_6, b_7$, $b_8$, and $N_7$ are given in the Appendix in  Eqs. (A1-A5).
The  post-evolution reduced density matrices of Bob and Charu are given by
\begin{widetext}
\begin{eqnarray}
\rho^{'U}_{B_G}= \frac{1}{4 N_7} \left(
\begin{array}{cc}
4  b_5 \cos ^2 \theta  \cos ^2\frac{x}{2}  & e^{-i y} \sin x ((b_5+b_6) \cos (2 \theta )+b_5-b_6) \\
e^{i y}  \sin x ((b_5+b_6) \cos (2 \theta )+b_5-b_6) & 4 b_5 \cos ^2 \theta  \sin ^2 \frac{x}{2}+b_6 \sin ^2  \theta  \cos ^2 \frac{x}{2} \\
\end{array}
\right)= \rho^{'U}_{C_G}.
\end{eqnarray}
\end{widetext}
 Comparing $\rho^{'U}_{BC}$ and $\rho'_{BC}$, we find that the no-signaling principle is satisfied ($\rho^{U}_{BC_{G}}  = \rho_{BC_{G}} $) if and only if $t_1 = 0$ or $\alpha = 0 $, which converts the $PT$-symmetric Hamiltonian to a Hermitian one. Hence again for $\rho'_{{ABC}_G}$, we can claim that when Bob and Charu are at the same location, the no-signaling principle is violated if the local operation on Alice's part is implemented by resorting to a $PT$-symmetric non-hermitian Hamiltonian with real eigenvalues.

Now, if Bob and Charu are at separate locations, we compare the post-evolution reduced density matrices of Bob and Charu, viz.   $\rho^{'U}_{B_{G}}$ and $\rho^{'U}_{C_{G}}$ respectively, with the initial ones, viz.   $\rho'_{B_{G}}$  and $\rho'_{C_{G}}$.
We find that the no-signaling principle is violated except for the surface of the parameter space given by $y=0$ and $x=\xi$. On that surface, 
we have
\begin{eqnarray} \nonumber
 \rho^{'U}_{B} &=& \rho'_{B} \\ \nonumber &=& \frac{1}{2}\left(
\begin{array}{cc}
  \cos (2 \theta ) \cos \xi +1 &  \cos (2 \theta ) \sin \xi  \\
  \cos (2 \theta ) \sin \xi  &  1-\cos (2 \theta ) \cos \xi  \\
\end{array}
\right)\\  & = & \rho^{'U}_{C} = \rho'_{C}.
\end{eqnarray}

In order to find the the best success probability of discriminating between $\rho^{'U}_{BC}$ and $\rho'_{BC}$, we need the four  eigenvalues of $\rho_{d_{3}} = (\rho^{'U}_{BC} - \rho'_{BC})$, which are obtained as $\lambda'_1 = 0$, $\lambda'_2 = 0$,
\begin{eqnarray}
\nonumber
\lambda'_3 =  -\frac{1}{N_8}\sin  \alpha  \cos ^2 \alpha  \sin t_1 \bigg(\cos (2 \alpha )\\ \nonumber -2 \sin ^2\alpha  \cos (2 t_1)\bigg)^{1/2},
\end{eqnarray}
and $\lambda'_4 = -\lambda'_3$, 
where $N_8 =\cos ^2 \alpha  \left(2 \sin ^2 \alpha  \cos (2 t_1)-2\right) $.
The best success probability for discriminating between the mixed states $\rho'_{BC}$ and $\rho^{'U}_{BC}$ is given by
\begin{equation}
P_{GHZ'} = \frac{1}{2}\left(1+ \left|\lambda'_3\right|\right).
\end{equation}
%
The highest value of the best success probability is approximately $0.75$ obtained at $t_1 \approx \pi/2$, $\xi \approx \pi/3$, and $\alpha \approx  \frac{9 \pi}{20}$. 

\section{Conclusion}
\label{pienc}
The \(PT\)-symmetric dynamics can be seen as going beyond the unitary dynamics of quantum mechanics 
while still retaining the standard static and measurement descriptions within the Hilbert space description of physical systems using the Dirac inner product between state vectors. 
It is known that the formalism can violate the no-signaling principle by using entangled state vectors of bipartite physical systems. It has already been noticed before that the no-signaling principle can be restored by using a non-conventional inner product. We showed that the use of multiparty entangled states can give rise to a situation where, even without wandering beyond the standard Dirac inner product of Hilbert spaces, the no-signaling principle is recovered at the local level of the different parties of the multiparty state, while the global system still violates the principle. We find that while the generalized Greenberger-Horne-Zeilinger states do support such a local preservation of no-signaling, the generalized W states do not.


\begin{widetext}

\appendix
\section{}

The following expressions  were used in the text.
\begin{eqnarray}
b_5 &&= \frac{1}{N_7}\bigg[\sec ^2 \alpha +\tan \alpha  (\sin (2 t_1) (\cos \xi  \sin x \cos y-\sin \xi  \cos x)\\ \nonumber &&-\tan \alpha  \cos (2 t_1))-2 \tan \alpha  \sec \alpha  \sin ^2 t_1 \sin x \sin y\bigg],\\
b_6 &&= \frac{1}{N_7}\bigg[ \sec ^2 \alpha  \bigg(\cos ^2 \alpha  \cos (2 t_1)+\sin \alpha  \cos \alpha  \sin (2 t_1) (\sin \xi  \cos x-\cos \xi  \sin x \cos y)\\ \nonumber &&+2 \sin ^2 t_1 (\sin \alpha  \sin x \sin y+1)\bigg)\bigg],\\
b_7 &&= \frac{1}{ N_7}\bigg[\tan \alpha  \left(2 \sec \alpha  \sin ^2 t_1 \cos x \sin y-\sin (2 t_1) (\sin \xi  \sin x+\cos \xi  \cos x \cos y)\right)\\ \nonumber &&-i \tan \alpha  \left(2 \sec \alpha  \sin ^2 t_1 \cos y+\cos \xi  \sin (2 t_1) \sin y\right)\bigg],\\
b_8 &&=\frac{1}{N_7}\bigg[\tan \alpha  \left(2 \sec \alpha  \sin ^2 t_1 \cos x \sin y-\sin (2 t_1) (\sin \xi  \sin x+\cos \xi  \cos x \cos y)\right)\\ \nonumber &&+i \tan \alpha  \left(2 \sec \alpha  \sin ^2 t_1 \cos y+\cos \xi  \sin (2 t_1) \sin y\right) \bigg],\\
N_7 &&=  \cos ^2 \theta  \left(\sec ^2 \alpha -\tan ^2 \alpha  \cos (2 t_1)\right)+\sin ^2 \theta  \left(2 \sec ^2 \alpha  \sin ^2 t_1+\cos (2 t_1)\right)\\ \nonumber && +\tan \alpha  \cos (2 \theta ) \left(-\sin \xi  \sin (2 t_1) \cos x -2 \sec \alpha  \sin ^2 t_1 \sin x \sin y+\cos \xi  \sin (2 t_1) \sin x \cos y\right).
\end{eqnarray}

\end{widetext}
\end{document}